\documentclass[12pt]{article} 

\usepackage{hyperref} 

\DeclareFixedFont{\xiiss}{OT1}{cmss}{m}{n}{12}
\DeclareFixedFont{\ixss}{OT1}{cmss}{m}{n}{9}
\DeclareFixedFont{\cmrnine}{OT1}{cmr}{m}{n}{9}
 
\newcommand{\CC}{\hbox{\xiiss C\kern-.4emI}}
\newcommand{\RR}{\hbox{\xiiss R\kern-.45emI}}\newcommand{\BR}{\RR}
\newcommand{\ZZ}{\hbox{\xiiss Z\kern-.4emZ}}\newcommand{\BZ}{\ZZ}
\newcommand{\CCs}{\hbox{\ixss C\kern-.4emI}}
\newcommand{\ZZs}{\hbox{\ixss Z\kern-.4emZ}}
\newcommand{\BCP}{\hbox{\xiiss C\kern-.4emIP}}

\newcommand{\beq}{\begin{equation}}
\newcommand{\beql}[1]{\begin{equation}\label{eq:#1}}
\newcommand{\eeq}{\end{equation}}
\newcommand{\be}{\begin{equation}}
\newcommand{\ee}{\end{equation}}
\newcommand{\beqn}{\begin{eqnarray}}
\newcommand{\eeqn}{\end{eqnarray}}
\newcommand{\bea}{\begin{eqnarray}}
\newcommand{\eea}{\end{eqnarray}}

\newcommand{\eq}[1]{(\ref{eq:#1})}

\begin{document}
        \begin{titlepage}
        \title{
                \begin{flushright}
                \begin{small}
                ILL-(TH)-99-01\\
                hep-th/9904040\\
                \end{small}
                \end{flushright}
               \vspace{1cm}
			Spacetime supersymmetry in $AdS_3$ backgrounds
		}

\author{ 	David Berenstein\thanks{email:{\tt berenste@hepux0.hep.uiuc.edu}}
			\\
			and \\
			Robert G. Leigh\thanks{email:{\tt rgleigh@uiuc.edu }}\\ 
        	\\
                {\small\it Department of Physics}\\
                {\small\it University of Illinois at Urbana-Champaign}\\
                {\small\it Urbana, IL 61801}\\
		}

        \maketitle

        \begin{abstract}
We construct string target spacetimes with $AdS_3\times X$ geometry,
which have an $N=2$ spacetime superconformal algebra. $X$ is found to be
a $U(1)$ fibration over a manifold which is a target for an $N=2$
worldsheet conformal field theory. We emphasize theories with free field
realizations where in principle it is possible to compute the full
one-particle string spectrum.
      \end{abstract}

        \end{titlepage}


\section{Introduction}

The AdS/CFT correspondence\cite{Mald} relates string theories and
M-theory  solutions with an $AdS$ factor to the infrared conformal field
theory living on the brane system. This was further elaborated upon in
Refs. \cite{Wi,GKP} where it was explained how one calculates the
spectrum and correlation functions of the conformal field theory within
a supergravity description. Tests of the conjecture involve comparing
the supergravity correlation functions and dimensions of operators with
the ones associated to the low energy dynamics of the brane.

In order to construct tests that probe beyond BPS states and the large
$g^2N$ limit, it is necessary to understand the full spectrum of
excitations in these backgrounds. The first example\footnote{For earlier
work, see Refs. \cite{morerefs}.} was discussed in
\cite{GKS} where a weak coupling regime exists at finite $N$ (see also
\cite{BORT,newKS}).  Further examples have appeared in Refs.
\cite{KLL,EFGT}. In general, in order to probe physics where
supersymmetry is less restrictive but still a powerful tool, we want to
develop a large collection of spacetimes that are good string
backgrounds and that can be understood in a simple fashion. It is the
purpose of this paper to give a
large class of backgrounds for which this program is possible in
principle.
Our construction gives rise to spaces with $N=2$ spacetime
supersymmetry, and they are derived from 
a worldsheet $N=2$ superconformal field theory.

After we completed this paper, we learned of the results of Ref. \cite{GR}
which overlaps some of our results.

\section{$AdS_3$ backgrounds}

%
We wish to discuss general $NSNS$ backgrounds of Type II superstrings
which lead to spacetime supersymmetry on $AdS_3$. We will require that 
a large radius limit of the geometry exists, such that a semiclassical 
analysis can be made valid; this is done in order to be able to 
compare with low energy supergravity.
We will restrict ourselves to coset conformal field theories, for which
the semiclassical limit corresponds to taking all levels of the
corresponding algebras to infinity simultaneously.
Moreover, we will be interested in theories with
$N=2$ worldsheet superconformal invariance, which will leads naturally
to spacetime supersymmetry.

These two requirements lead to the following constraints.
The spacetime is of the form $AdS_3 \times X$, with $X$ a 
seven-dimensional manifold/orbifold. $\hbox{N=2}$ worldsheet supersymmetry is
obtained if $X$ is chosen to be a circle bundle over a complex manifold,
which itself is a product of hermitian symmetric spaces constructed as in
Kazama-Suzuki\cite{KZ}. (A possible generalization would be to take
more general complex manifolds $X/U(1)$
with constant positive curvature, but this is less well understood.)

This bundle structure allows for a decomposition of the energy momentum
tensor in terms of a $U(1)$ CFT and a
hermitian symmetric space coset. This $U(1)$
pairs with the Cartan of the $SL(2)$ current algebra into an $N=2$ free
system, and the remaining $SL(2,\BR)/U(1)$ is also a hermitian symmetric
non-compact coset. This splitting of the $U(1)$  piece of $X$ is required if
we want to bosonize the fermions in the $SL(2,\BR)$ current algebra. 

In total, we have a system with $N=2$ worldsheet supersymmetry,
\beq
\left( SL(2,\RR)\slash U(1)\right)\times
\left( X\slash U(1)\right)\times U(1)^2
\eeq
This construction possesses a ``natural'' GSO projection (preserving the
$N=2$ structure) which leads to spacetime supersymmetry. Of course,
there are other possible GSO projections, such as that used in the
construction of the NS vacua of the $N=4$ geometries considered in \cite{GKS,KLL,EFGT}. 
Each of these GSO projections will be considered below.

The collection of models constructed in this fashion is a finite family
of spaces, and they should be interpreted as conifolds where a set of
D1-D5 branes is localized.

A table with a list of compact cosets of dimension $d\leq 7$ is
provided here, where we include the central charge of the model in terms
of the levels of the different groups that form the coset.
Other spaces like $SU(3)/U(1)^2$ can be constructed as $SU(3)/(SU(2)\times U(1))\times
SU(2)/U(1)$, so we don't list them. 

\renewcommand{\arraystretch}{1.5}
\begin{table}[h]
\begin{tabular}{c|c|c|c}
Coset & Central charge & Group& Dimension\\
\hline
$S^1$& $\frac 32$ & U(1) & 1\\
$\BCP^1$ & $3-\frac 6 k$ & $SU(2)_k\slash U(1)$ & 2\\
$S^3$&$\frac 92 -\frac 6k$ & $SU(2)_k$&3\\
$\BCP^2$ & $6-\frac {18} k$ & $SU(3)_k/(SU(2)_k\times U(1))$ & 4\\
$S^5$ & $\frac{15}2 - \frac {18} k$ & $SU(3)_k/SU(2)_k$ & 5 \\
$T^{pq}$ & $\frac {15}2 - \frac 6k -\frac 6{k'}$ & $(SU(2)_k\times SU(2)_{k'})/U(1)$&5\\
$\BCP^3$ & $9 -\frac {36}k$ & $SU(4)_k/(SU(3)_k\times U(1))$ & 6\\
$S^7$ & $\frac{21}2 -\frac {36}k$ & $SU(4)_k/SU(3)_k$&7\\
$T^{pqr}$ & $\frac {21}2 - \frac 6 k -\frac 6{k'}-\frac 6{k''}$&
$(SU(2)_k\times SU(2)_{k'}\times SU(2)_{k''})/U(1)^2$&7
\end{tabular}
\end{table}

Our objective in the next section will be to give a detailed
construction of spacetime supersymmetry in these backgrounds, and to
show why the standard choice of using the $N=2$ $U(1)_R$ worldsheet
current is not appropriate. 

We begin by defining our conventions. In general, we will have an
$SL(2,\BR)_k$ superconformal current algebra $(J^a,\psi^a)$, realized by
free fields. 
There is a free $U(1)$ system, with currents $(K,\chi)$. Finally, there
is a coset $X/U(1)$ with an $N=2$ $U(1)_R$ current which we will denote
by $J_c$. We require the total central charge $c_T=15$ such that it may be coupled 
to the standard superconformal ghost system to define a consistent string theory.
 
The total $U(1)_R$ current may then be written as
\newcommand{\jr}{J_R}
\begin{equation}
\jr = (\psi^0 \chi)+\left( :\psi^+\psi^-: 
+\frac{2}{k} J^0  + J_c \right)= 
i\partial H_1 +i\partial H_2
\end{equation}
We have bosonized the currents in parenthesis in favor of two 
bosons\footnote{Note the normalization $H_2(z)H_2(0)=-4\ln z$.}
$H_1,H_2$. This is the canonical choice for generic $X$, where we require
the bosons to be integral.\footnote{In special cases,
$X$ will be such that $\jr$ splits integrally in terms of more bosons. In
these cases, spacetime supersymmetry will be enhanced.}

It is straightforward to verify that the following OPEs are
regular
 \beqn
 \jr(z) J^0(z') \sim 0\\
 \jr(z) K(z') \sim 0
 \eeqn
 and moreover
 \begin{eqnarray}
 \jr(z) \jr(z')&\sim& \frac {5}{(z-z')^2}\label{eq:RRope}\\
 J^0(z) J^0(z') &\sim & -{k/2\over (z-z')^2}\\
 K(z) K(z') &\sim & \frac {k'}{(z-z')^2}
 \end{eqnarray}
where $k$ is the level of the $SL(2)$ current algebra and $k'$ is
related to the radius of compactification of the free boson leading to
the $U(1)$ supercurrent. In deriving \eq{RRope}, we have used $c=15$
(and thus $c(X/U(1))=9-6/k$).

The canonical choice for spacetime supercharges is
\beql{standQ}
Q^{\pm\pm} = \oint dz\ e^{\pm iH_1/2\pm iH_2/2} e^{-\phi/2}
\end{equation}
where $\phi$ is the superconformal ghost. These supercharges have the
property that they are BRST invariant, while mutual locality requires
that we keep only $Q^{+\pm}$. Modulo picture changing\cite{FMS}, we find
the commutator
\begin{equation}
\{Q^{++}, Q^{+-}\} = J^0 -K
\end{equation}
Thus, \eq{standQ} leads naturally to spacetime supersymmetry, but
unfortunately, the expected spacetime bosonic symmetries (such as
$SL(2)$) are not recovered. (problems of this nature were discussed in
\cite{GKS}; there the problems were much worse, as the algebra was
doubled, etc.) We'll consider this in more detail below, and give the
correct construction.

\section{Target space consistency}

If $X/U(1)$ is a direct product $\sim \oplus_i {\cal A}_i$, where each
factor has central charge $c_i=b_i-a_i/k_i$, then the condition $c=15$ 
forces the constraint
\begin{equation}\label{eq:sevencon}
\frac 6 k = \sum_i \frac{a_i}{k_i}
\end{equation}
which relates the level of the $SL(2)$ algebra to the level of the other
cosets in the construction, and it is easy to see that taking $k_i\to
\infty$ for all $i$ forces $k\to\infty$. This means that we have a good
geometric picture of the coset as a target space, since in the large
$k_i$ limit the cosets behave semiclassically. In writing \eq{sevencon},
we have assumed that $X$ has classical dimension seven. Changing this
would modify \eq{sevencon} by terms of order one, so the AdS spacetime
would never become semiclassical.

\newcommand{\js}{{\cal J}}
Now let us analyse the spacetime isometries. We would like to see that
the isometries have appropriate commutators with the spacetime
supercharges. In particular, given the worldsheet $SL(2)$ currents, we
can construct charges that should generate spacetime symmetries;
consider the spacetime $SL(2)$ raising operator $\js^+ = \oint J^+ dz$.
This operator has the following OPE with $\jr$,
\begin{equation}
\js^+ \jr(z) \sim -\psi^+ \chi +2\psi_0\psi^+ - \frac {2}{k} J^+
\end{equation}
and the last term implies that the OPE $\js^+ Q^{++}$ is
non-local.\footnote{To see this, it is useful to realize $Q\sim e^{\int
\jr}$.}  Also, the $Q$'s do not carry $U(1)_R$ charges, which is in
conflict with the spacetime $N=2$ algebra if we want to identify $\oint
K$ as the $U(1)_R$ current. That is, the existence of the operator
$Q^{++}$ leads to the wrong target space symmetries.

Instead of doing this, we will modify the supersymmetry generator in
order that the spacetime Virasoro algebra is recovered.

It is simple to see that the problem comes from the factor of $J^0/{2k}$
in the $U(1)_R$ current of $SL(2,\BR)/U(1)$. To solve the problem then,
it is sufficient to redefine the worldsheet $U(1)_R$ current
\begin{equation}
\tilde\jr = \jr +\frac 1{2k}J^0 = i\partial H_1+i\partial \tilde H_2
\end{equation}
However, the OPE $\tilde\jr(z)\tilde\jr(z') \sim (5 - \frac 1{4k})/(z-z')^2$ 
implies that the spin operator
\begin{equation}
\Sigma^{++} = \exp(+\frac {iH_1}{2}+\frac{i\tilde H_2}{2})
\end{equation}
has the wrong ($\neq 5/8$) conformal dimension. This means we need to
correct again $\tilde\jr$. That is, we define
\begin{equation}
\jr' = \tilde\jr + \frac F{2k} 
\end{equation}
with $F$ a current such that $F(z)F(z') = - J^0(z) J^0(z')$ and that is
orthogonal to $J_0$, so that the bosonization of $J'$ leads to the
correct conformal weight for the spin operator. We have a canonical
choice for $F$ in our models, as we have the extra free $U(1)$, $K$.

Hence, if $K$ has the same level as $SL(2)$ , then
\begin{equation}
\jr' = \jr + \frac{ J^0+K}{2k}
\end{equation}
will lead to the correct conformal weight for the spin operator. Notice
that the current $J^0+K$ is null, and orthogonal to $\jr$, so that the
natural spin operators constructed from $\jr'$ have automatically the
correct conformal dimension. This also gives a charge to the spin
operators under the spacetime $U(1)_R$ generated by $\oint K$.

The full current is  a sum of three integral  pieces now
\begin{equation}
\jr' = (\psi_0 \chi) + (\psi^+\psi^-) + (\frac {K}{2k} + J_c) 
= i\partial(H'_1+H_2'+H'_3) 
\end{equation}
Let us check that we get the $N=2$ superconformal algebra in spacetime.

The BRST invariant spin operators will be given by
\begin{equation}
S = \exp(\epsilon_i\frac{iH'_i}2)\exp(-\phi/2)
\end{equation}
with $\epsilon_i= \pm 1$. Mutual locality and BRS invariance lead to
the constraint $\prod \epsilon_i = 1$, and the spacetime
supersymmetry operators are given by
\begin{eqnarray}
Q^{+\pm} &=& \oint \exp\left({i\frac {H'_1}{2} 
\pm i\frac {H'_2}{2}\pm i\frac {H'_3}{2}}\right)\\
Q^{-\pm} &=& \oint \exp\left({-i\frac {H'_1}{2} 
\pm i\frac {H'_2}{2}-\pm i\frac {H'_3}{2}}\right)
\end{eqnarray} 
The spacetime supersymmetry algebra (modulo picture changing) is now
seen to be given by
\begin{eqnarray}
\{Q^{++},Q^{-+}\} &=& J^+\\
\{Q^{+-},Q^{--}\} &=& J^-\\
\{Q^{-+},Q^{--}\} &=& J^0+K\\
\{Q^{+-},Q^{++}\} &=& J^0 - K
\end{eqnarray}
which is exactly the $NS$ sector of two dimensional $N=2$ supersymmetry.
That is, this provides a general construction of $N=2$ spacetime
supersymmetry whenever we have $N=2$ worldsheet  supersymmetry. Motions
in the $U(1)$ fiber are the generators of the $R$ symmetry spacetime
current, and the modifications we made give charge to the supersymmetry
generators, which is a requirement of the algebra.

We want to  complete the construction with the rest of the Virasoro
algebra in spacetime. That is, we can write the rest of the modes of the
spacetime supercurrent by applying the other Virasoro generators to the
spin fields \cite{GKS}.

It is a straightforward exercise to show that the algebra closes with a
central charge given by $c_{st} = k p$, where $p$ is the winding number
of the strings located at infinity.

It is also clear that the fact that the level of the $U(1)$ is chosen to
be equal to that of the $SL(2)$ current implies a quantization condition
on the radius of the $U(1)$ bundle to be of the same order of magnitude
as the radius of the AdS spacetime.

\section{Comments on $AdS_3\times S_3\times T^4$ and 
$AdS_3\times S^3\times S^3\times U(1)$}

Now that we have a general recipe, we can go back and analyze previous
results \cite{GKS, EFGT} in a new light. We confine ourselves to a few
comments.

In particular, the current $K$ is chosen in a very special way. For
example, in $AdS_3\times S^3\times T^4$, we could have chosen either the
Cartan of $SU(2)$, or one of the free $U(1)$'s. Since we have only analyzed
the effects of the spacetime raising operator $J^+$, we can always
guarantee that we get the spacetime Virasoro algebra with either choice.
On the other hand, if we want to keep the full $SU(2)$ isometries of the
target space, we actually have to also eliminate a similar problem with
the locality of the $SU(2)$ raising operator, and the modification in
the currents is such that both effects get cancelled simultaneously if
we choose to pair the Cartan element of $SU(2)$ with the one of
$SL(2,\BR)$, so the choice of null field solves both problems at the
same time.

For $AdS_3\times S^3\times S^3\times U(1)$, there is one linear
combination which we want to add to the Cartan of $SL(2,\BR)$ such that
all of the problems with the isometries go away simultaneously. This
choice is the diagonal $U(1)$ which cancels the terms from both $AdS_3$
terms, and a posteriori ends up having the correct quantization
condition. This chooses a complex structure on the four free $U(1)$
directions.

Finally, let us remark on $AdS_3\times S^3\times \BCP^2$, where $\BCP^2$
is understood as the coset $SU(3)/(SU(2)\times U(1))$. In this case, it
can be seen that the Cartan of the $SU(2)$ coming from $S^3$ can not
cancel the non-locality of the raising operator of $SU(2)$, so that we
do not preserve the isometries, and the correct quantization condition
squashes the sphere, so it leaves us with $N=2$ spacetime supersymmetry
as opposed to $N=4$ which one may have naively guessed.

\section{Vertex Operators and free fields.}

We have shown that it is straightforward to compute the spacetime  NS
sector of supersymmetry. Naturally, as we have a field of conformal
dimension zero, represented by $\gamma$, we can multiply $Q$ by any
fractional powers of $\gamma$. BRST invariance will choose a linear
combination of $Q\gamma^\alpha$ with the same spacetime quantum numbers.

Closure of the fermionic algebra on the bosonic generators provides that
$\alpha\in \BZ$ or $\alpha\in \BZ+1/2$, but not both simultaneously, as
then the bosonic generators will not be single valued when $\gamma\to
e^{2\pi i}\gamma$. That is, we have a choice of two super selection
sectors for the supersymmetry algebra. One is to be taken as the
Neveu-Schwarz sector as we already saw, and the other is the Ramond
sector of the spacetime supersymmetry.

This is a feature of the free field realization for the $SL(2)$ current
algebra, and it actually lets us write vertex operators for the $SL(2)$
current with free fields. The same can be done with the coset
constructions, as they admit free field realizations. Now, we want to
analyze the chiral GSO projection. This needs to be done for each case,
as we are not using the $N=2$ structure of the worldsheet CFT directly,
but a modified version of it.

Let us fix some notation.

As is well known, coset conformal field theories admit free field
realizations \cite{KOS,KOS2}. For each of the raising operators of the
algebra we use a $B,C$ system, with OPE given by
\begin{equation}
B(z) C(z') = \frac 1{z-z'-\theta\theta'} = \frac{1}{Z-Z'}
\end{equation}
these can be super-bosonized into a set of two null (lightcone) scalars
$T,U$ with OPE
\begin{equation}
T(z) U(z') = \log(Z-Z')
\end{equation}
by taking $B = a DU \exp (a^{-1}T)$, $C = \exp (-a^{-1}T)$.

To the Cartan elements we associate free supersymmetric bosons $H$. The
total collection of free bosons will be labeled by $\phi$, and the
fermions by $\psi$.

For the bosons we want to take operators (in the left moving sector lets
say) which have an odd number of free fermion insertions in the $[-1]$
picture.
\begin{equation}
 {\cal O}(\psi^\alpha, \partial\phi, \partial)e^{ip \cdot \phi}
\end{equation}
In this basis the embedding of the group theory (allowed lattice of
values for $p$) is not manifest.

This requirement corresponds to the chiral GSO projection. The mass
shell condition
\begin{equation}
p\cdot (p-\lambda) +N = \frac 12
\eeq
may be equated to $-\frac {j(j+1)}{2k} + m^2 +1/2$, where $j$ is the
spacetime $SL(2)$ quantum number and $m^2$ is to be understood as the
$AdS_3$ mass of the state. $\lambda$ is the curvature coupling of the
free scalars. The GSO projection implies that we get  a positive value
for $j$; that is, the theory does not have tachyons.

The number of free superfields is ten, and the total field with a
worldsheet curvature coupling of the $AdS$ and $X$ theories combined
together is a null field. In this sense, all theories have the same
underlying structure. It is the choice of lattice (modular invariant)
which makes them different from one another.

The ten dimensional massless vertex operators are  the most interesting
as they predict the supergravity spectrum of the compactification. In
this case the polynomial $S$ reduces to one free fermion operator.
Amongst these, we will find all the chiral operators of the spacetime
conformal field theory which can be described by string vertex
operators.

The fermion vertex operators are constructed by acting with the
spacetime supersymmetry generators on the spacetime boson vertex
operators that we described. As we have a choice of super selection
sector, they will look different in each of the cases. It is also clear
that this difference is in the powers of $\gamma$, and therefore the OPE
of two fermion vertex operators closes on  the ones of bosonic type.

The advantage of having a free field realization is that we have a
choice of ten free fermions $\psi^i$, which can be bosonized into five
scalars, and writing spin vertex operators is straightforward. One has
to remember that in order to get the right charges for the
supersymmetries, they will be multiplied by powers of $e^{ip\cdot\phi}$
with $p^2 = 0$.

From the bosonization, we find that each $B,C$ system
contributes two lightcone scalars, i.e., their signature is
$(1,1)$. Hence the lattice that we obtain has a signature of $(n,m)$
with $n+m=10$, and $n,m\geq 2$. The difficulty in writing the partition
function lies in finding the constraints on the lattice and the
screening operators of the system, so that at the end we can recover
unitarity and modular invariance.

\section{Discussion}

We have given a  complete construction of $N=2$ spacetime supersymmetry
on $AdS_3$ spacetime for type II NS NS backgrounds, which are
constrained to admit a version of $N=2$ supersymmetry on the worldsheet.
It is clear that as we have exact conformal field theories on the
worldsheet, these are solutions to all orders in $\alpha'$ of the string
equations of motion. Moreover, usually when we have enough spacetime
supersymmetry, this is what we need to guarantee that there are no
non-perturbative corrections to the target space. In principle, if we
ask questions that can be answered perturbatively, this approach should
give a complete set of calculational tools.

On the other hand, further work is required. In particular, the coset
model predictions certainly go beyond supergravity and compute the full
spectrum of the one-particle states of the string theory propagating on
these spacetimes, and therefore, we can get a good idea of what the
spacetime conformal field theory might be, even when we are not required
to be in any large radius limit. This should  shed light on the $\frac
1N$ corrections to the conformal field theories.

The detailed form of the string partition function is not known, and it
is of course important to construct it and check modular invariance.
This is not clear, as we can not go to a lightcone gauge where
everything is manifestly unitary. Also, as we have constructed these
theories with free fields, it is likely that everything is complicated
by the screening charges, so that a very good knowledge of how to
extract the real physical degrees of freedom is required. Although some
progress has been made in \cite{DQ}, it is far from complete.

Spaces with non-compact cosets \cite{GQ,BN} are also interesting, but
their spacetime CFT description is bound to be complicated by the non
compactness, as the meaning of the conformal boundary  comes into
question, and if we get a spacetime CFT it seems naturally to give rise
to a continous spectrum  of states which is certainly more difficult to
analyze.

\medskip

\noindent {\bf Acknowledgments:} We wish to thank F. Larsen for
discussions. Work supported in part by the United States Department of
Energy grant DE-FG02-91ER40677 and an Outstanding Junior Investigator
Award. 

%
%
%

\providecommand{\href}[2]{#2}\begingroup\raggedright
\endgroup

\end{document}